\documentclass[10pt]{emulateapj}
\usepackage{natbib}

\shorttitle{Solar Wind Turbulent Heating}
\shortauthors{Howes}

\newcommand\Alfven{Alfv\'en }
\newcommand\Alfvenic{Alfv\'enic }

\newcommand{\figref}[1]{Figure~\ref{#1}}
\newcommand{\secref}[1]{\S\ref{#1}}
\newcommand{\eqref}[1]{equation~(\ref{#1})}

\begin{document}


\title{Prediction of the Proton-to-Total Turbulent Heating in the Solar Wind}


\author{G.~G. Howes}
\affil{Department of Physics and Astronomy, University of
Iowa, Iowa City, IA, 52242}

\begin{abstract}
This paper employs a recent turbulent heating prescription to predict
the ratio of proton-to-total heating due to the kinetic dissipation of
Alfv\'enic turbulence as a function of heliocentric distance. Comparing
to a recent empirical estimate for this turbulent heating ratio in the
high-speed solar wind, the prediction shows good agreement with the
empirical estimate for $R \gtrsim 0.8$~AU, but predicts less ion
heating than the empirical estimate at smaller heliocentric radii. At
these smaller radii, the turbulent heating prescription, calculated in
the gyrokinetic limit, fails because the turbulent cascade is
predicted to reach the proton cyclotron frequency before Landau
damping terminates the cascade.  These findings suggest that the
turbulent cascade can reach the proton cyclotron frequency at $R
\lesssim 0.8$~AU, leading to a higher level of proton heating than
predicted by the turbulent heating prescription in the gyrokinetic
limit.  At larger heliocentric radii, $R \gtrsim 0.8$~AU, this
turbulent heating prescription contains all of the necessary physical
mechanisms needed to reproduce the empirically estimated
proton-to-total heating ratio.
\end{abstract}


\section{Introduction}
In the development of a thermodynamic model of the heliosphere, a
crucial issue is the identification of the various physical mechanisms
that play a role in determining the temperature profiles of the
heliospheric plasma ions and electrons. Some of the effects
influencing the measured temperature profiles in the inner heliosphere
are energy conservation in the spherically expanding plasma, heat
conduction, Coulomb collisions, and plasma heating through the
dissipation of solar wind turbulence.\footnote{Pickup ions are
believed to significantly affect the heliospheric energy balance only
in the outer heliosphere at $R
\gtrsim 10$~AU \citep{Breech:2009}.}
The incorporation of the first three of these effects is relatively
well understood, but the heating of the heliospheric ions and electrons
resulting from the dissipation of solar wind turbulence remains an
area of active research.

Although a number of studies have addressed the issue of plasma
heating by the dissipation of turbulence (see
\citet{Cranmer:2009} for a review of previous efforts), with the exception 
of a pioneering series  of papers by Quataert and Gruzinov
\citep{Quataert:1998,Gruzinov:1998,Quataert:1999}, 
the turbulent heating of the ions and electrons separately has only
recently become a focus of interest.  We review here three studies
that have made progress in the investigation of turbulent plasma
heating and the effect of this heating on the solar wind proton and
electron temperature profiles.

\citet{Breech:2009} coupled an existing  turbulence transport model for the
solar wind \citep{Zank:1996,Matthaeus:1999,Smith:2001a,Breech:2008} to
separate radial evolution equations for the proton and electron
temperatures. The temperature equations included the effects of
spherical expansion, parallel electron heat flux, collisional energy
exchange between protons and electrons, and separate turbulent heating
rates for the ions and electrons. The kinetic dissipation mechanisms
that determine the partitioning of turbulent heating between protons
and electrons were not addressed in this model, so the fraction of
proton-to-total turbulent heating was set to a constant value, $f_p=
Q_p/(Q_p+Q_e)$. The model achieved a reasonable accounting for the
temperature data measured by \emph{Ulysses} when the effect of Coulomb
collisions was weak, the electron heat conduction was provided by an
empirically determined function, and the partitioning of turbulent
heating was taken as $f_p=0.6$. Note that the value of $f_p=0.6$ is
consistent with other estimates of the fraction of proton heating
\citep{Leamon:1999,Stawarz:2009}.

In a complementary work, \citet{Cranmer:2009} constructed a model
following the internal energy evolution of the protons and electrons in
the solar wind in an attempt to constrain empirically the required
proton and electron turbulent heating in the solar wind. Assuming
isotropic Maxwellian proton and electron velocity distributions with
shared bulk velocity, separate equations for the conservation of
energy in a spherically expanding flow for protons and electrons were
constructed to incorporate the effects of energy exchange by Coulomb
collisions, parallel electron heat conduction, and turbulent heating
of protons and electrons. Using analytic fits to \emph{Helios} and
\emph{Ulysses} measurements for the proton temperature, electron 
temperature, and parallel electron heat conduction flux, the equations
were solved for the turbulent proton and electron heating
rates. The results were then combined to obtain an empirical estimate
of the proton-to-total heating rate $Q_p/(Q_p+Q_e)$ as a function of
heliocentric radius, reproduced as the dashed line in \figref{fig:qpqtot}.

Based on a theoretical model of the turbulent cascade of energy in a
weakly collisional plasma \citep{Howes:2008b}, \citet{Howes:2010d}
predicted the proton-to-electron heating ratio $Q_p/Q_e$ resulting
from the dissipation of Alfv\'enic turbulence. The key result of this
study was an analytical prescription for the heating ratio
$Q_p/Q_e(\beta_p, T_p/T_e)$, a function of only two plasma parameters,
the proton plasma beta $\beta_p$ and the proton-to-electron
temperature ratio $T_p/T_e$. The limits of validity of this heating
ratio prediction were given as a constraint on the minimum scale of
turbulent energy injection (assuming an isotropic driving mechanism
for the turbulence).

This paper describes the application of the \citet{Howes:2010d}
heating prescription to predict the proton-to-total turbulent heating
rate $Q_p/(Q_p+Q_e)$ for the high-speed solar wind and compares the
resulting prediction to the empirical estimate of
\citet{Cranmer:2009}.

\section{Prediction of Turbulent Heating}
\label{sec:results}
This section describes the prediction of the ratio of the
proton-to-total turbulent heating $Q_p/(Q_p+Q_e)$ in the high-speed
solar wind as a function of heliocentric radius $R$ using the
turbulent heating prescription by \citet{Howes:2010d}. Since this
prescription depends on the plasma parameters $\beta_p$ and $T_p/T_e$,
it is necessary to construct a model of the high-speed solar wind to
determine the variation of these plasma parameters with heliocentric
radius. In \secref{sec:model}, we describe the solar wind model. In
\secref{sec:turbmodel}, we review the theoretical framework of  low-frequency,
anisotropic \Alfvenic turbulence in a magnetized, weakly collisional
plasma that underlies the turbulent heating prescription. This
prescription is employed to predict the proton-to-total turbulent
heating $Q_p/(Q_p+Q_e)$ in \secref{sec:prediction}. In
\secref{sec:width}, we estimate the evolution of the width of the
inertial range in the high-speed solar wind in order to verify the
validity of the turbulent heating prescription in
\secref{sec:limits}.

\subsection{Solar Wind Model}
\label{sec:model}
We adopt the same specific model for the high-speed
solar wind used by \citet{Cranmer:2009} to facilitate the comparison
to their empirical turbulent heating estimate.  This model is used to
specify, as a function of heliocentric radius $R$, the two key plasma
parameters required by the turbulent heating prescription: the proton
plasma beta $\beta_p$ and the proton-to-electron temperature ratio
$T_p/T_e$.

Analytic fits to \emph{in situ} measurements of the high-speed solar
wind (faster than 600~km s$^{-1}$) from the \emph{Helios} and
\emph{Ulysses} spacecraft over the range $0.29\mbox{ AU} <R<5.4\mbox{
AU}$ were used by
\citet{Cranmer:2009} to generate equations for the proton and electron
temperatures as function of heliocentric radius $R$,
\begin{equation}
\ln \left(\frac{T_p}{10^5\mbox{ K}} \right) = 0.9711 - 0.7988 x + 0.07062 x^2
\label{eq:tp}
\end{equation}
\begin{equation}
\ln \left(\frac{T_e}{10^5\mbox{ K}} \right) = 0.03460 - 0.4333 x + 0.08383 x^2,
\label{eq:te}
\end{equation}
where  $x \equiv \ln(R/1\mbox{ AU})$. 

In addition to the proton temperature, we need to specify the form of
the proton density and magnetic field strength to determine the proton
plasma beta $\beta_p= 8 \pi n_p T_p/B^2$. Following
\citet{Cranmer:2009}, we take a proton density of the form 
\begin{equation}
n_p(R) = n_0 (R/1\mbox{ AU})^{-2}
\label{eq:n}
\end{equation}
where $n_0 = 2.5\mbox{ cm}^{-3}$.  The empirical turbulent heating
constraints calculated by \citet{Cranmer:2009} used a colatitude
$\theta=15^\circ$ to model the high-latitude \emph{Ulysses}
measurements. For the heliocentric distances covered by this model,
the winding of the magnetic field into the Parker spiral for the
high-speed streams at this colatitude is relatively weak, so a simple
monopolar model for the magnetic field strength is a reasonable
approximation,
\begin{equation}
B(R) = B_0 (R/1\mbox{ AU})^{-2},
\label{eq:b}
\end{equation}
with $B_0 = 2.5 \times 10^{-5}\mbox{ G}$.

Using these functions for $T_p$, $T_e$, $n_p$, and $B$ in high-speed
solar wind streams, we find that the proton plasma beta varies from
$\beta_p=0.92$ at 0.29~AU to $\beta_p=34$ at 5.4~AU, and the
proton-to-electron temperature ratio varies from $T_p/T_e=3.9$ at
0.29~AU to $T_p/T_e=1.3$ at 5.4~AU.

\subsection{Turbulent Cascade Model}
\label{sec:turbmodel}
The heating prescription presented in \citet{Howes:2010d} is
determined using a model for the turbulent cascade of energy in a
magnetized, weakly collisional plasma \citep{Howes:2008b}. The cascade
model determines the steady state form of the magnetic energy spectrum
of \Alfvenic fluctuations, based on three primary assumptions: (1) the
Kolmogorov hypothesis that the energy cascade is determined by local
interactions
\citep{Kolmogorov:1941}; (2) the turbulence maintains a state of
critical balance at all scales \citep{Goldreich:1995}; and (3) the
linear kinetic damping rates are applicable in the nonlinearly
turbulent plasma. 

The dependence of the nonlinear energy transfer rate on the local
turbulent fluctuations in the cascade model \citep{Howes:2008b} is
inspired by the following theoretical picture of low-frequency,
anisotropic \Alfvenic turbulence in a magnetized, weakly collisional
plasma
\citep{Howes:2008c,Schekochihin:2009}. The energy of
\Alfvenic fluctuations is injected into the turbulence isotropically
at a scale much larger than the ion Larmor radius, $L_0 \gg \rho_i$,
corresponding to an isotropic driving wavenumber $k_0 \rho_i \ll
1$. Since the damping of \Alfvenic fluctuations at this large scale by
wave-particle interactions in a weakly collisional plasma is
negligible, the turbulent fluctuations rise to sufficient
amplitudes that nonlinear interactions between counter-propagating
\Alfven wave packets transfer the turbulent fluctuation
energy to smaller scales. This sets up a critically-balanced,
anisotropic cascade of MHD \Alfven waves over all scales down to the
perpendicular scale of the ion Larmor radius, $k_\perp \rho_i \lesssim 1$
\citep{Goldreich:1995,Boldyrev:2005}\footnote{Perpendicular and parallel are defined with respect to 
the direction of the local mean magnetic field.}. Even in a weakly collisional plasma,
the dynamics of this \Alfven wave cascade is rigorously described by
the equations of reduced MHD \citep{Schekochihin:2009}. At the
perpendicular scale of the ion Larmor radius $k_\perp \rho_i \sim 1$,
the turbulence transitions to a critically balanced, anisotropic
cascade of kinetic \Alfven waves over the perpendicular scales
$k_\perp \rho_i \gtrsim 1$. 

The range of scales traversed by the MHD \Alfven wave cascade, between
the driving scale and the ion Larmor radius scale, is commonly
designated the ``inertial range''
\citep{Kolmogorov:1941} of MHD turbulence---\emph{i.e.}, the range 
of scales over which the effects of driving and dissipation are
negligible. For a sufficiently large inertial range, the anisotropy of
the energy cascade ($k_\parallel \propto k_\perp^{2/3}$ in the
Goldreich-Sridhar theory, or $k_\parallel \propto k_\perp^{1/2}$ in
the Boldyrev theory) leads to turbulent fluctuations, at the
transition to the kinetic \Alfven wave cascade at $k_\perp \rho_i \sim
1$, that are highly elongated along the direction of the local
magnetic field, $k_\parallel /k_\perp \ll 1$.  Such 
anisotropic fluctuations are optimally described by a low-frequency
expansion of kinetic theory called gyrokinetics
\citep{Rutherford:1968,Frieman:1982,Howes:2006,Schekochihin:2009}. 
For \Alfvenic fluctuations, the anisotropy implies that, even at
scales $k_\perp \rho_i \sim 1$, the turbulent fluctuation frequency
remains much smaller than the ion cyclotron frequency, $\omega \ll
\Omega_i$. This is an important limit for the applicability of
gyrokinetic theory, and enables one to determine quantitatively the
limit of validity of the heating prescription (see
\secref{sec:limits}).  When this limit is satisfied, the ion cyclotron
resonance plays a negligible role in the collisionless damping of the
turbulent fluctuations \citep{Lehe:2009}.  Instead, collisionless
damping of the fluctuations occurs via the both the ion and electron
Landau resonances at scales $k_\perp
\rho_i \gtrsim 1$, implying that the kinetic \Alfven wave cascade comprises
the ``dissipation range'' of \Alfvenic turbulence.

It is in the dissipation range that wave-particle interactions
transfer the electromagnetic fluctuation energy to the ion and
electron particle distribution functions. Ultimately, this free energy
in the particle distribution functions is transferred to small scales
in velocity space through an entropy cascade
\citep{Schekochihin:2009, Tatsuno:2009,Plunk:2010,Plunk:2011}, enabling arbitrarily weak 
collisions to thermalize this energy, increasing the entropy and
leading to irreversible heating of the plasma. Wave-particle
interactions via the Landau resonance typically peak at $k_\perp
\rho_i \sim 1$ for ions and $k_\perp \rho_i > 1$ for electrons, so the 
predicted result for ion-to-electron heating $Q_i/Q_e$ is sensitive to
the model for the nonlinear energy transfer rate in the kinetic
\Alfven wave cascade.

Although the physical model of the turbulent cascade presented here
remains controversial within the heliospheric physics community, there
exists significant numerical and observational evidence in support of
two of its key features: (1) the turbulent frequency remains low,
$\omega \ll \Omega_i$, even for $k_\perp \rho_i \gtrsim 1$; and (2)
the turbulence transitions to a cascade of kinetic \Alfven waves at
$k_\perp \rho_i \sim 1$. A gyrokinetic numerical simulation of the
transition from the MHD \Alfven wave to the kinetic \Alfven wave
cascade at $k_\perp \rho_i \gtrsim 1$ \citep{Howes:2008a} produces
magnetic and electric energy spectra that are consistent with
\emph{Cluster} measurements of turbulence in the solar wind
\citep{Bale:2005}. A recent gyrokinetic simulation spanning the 
entire dissipation range from the ion to the electron Larmor radius
\citep{Howes:2011a} yields a magnetic energy spectrum that is quantitatively consistent
with \emph{in situ} measurements of the dissipation range turbulence
up to 100~Hz
\citep{Sahraoui:2009,Kiyani:2009,Alexandrova:2009,Chen:2010,Sahraoui:2010b}.
The striking agreement between the predictions for a kinetic \Alfven
wave cascade and the observed magnetic and electric power spectra
found by \citet{Sahraoui:2009} provide observational support for this
model. Finally, a $k$-filtering analysis of multi-spacecraft
\emph{Cluster} measurements demonstrates that the wavevectors of the
turbulent fluctuations at scales $k_\perp
\rho_i \sim 1$ are aligned nearly perpendicular to the local magnetic
field \citep{Sahraoui:2010b}; for \Alfvenic turbulent fluctuations,
this implies low turbulent frequencies, $\omega \ll \Omega_i$, in
support of the turbulent model employed in this study.

\subsection{Turbulent Heating Prediction}
\label{sec:prediction}
Assuming a fully ionized plasma of protons and electrons with
isotropic Maxwellian equilibrium velocity distributions,
\citet{Howes:2010d} used  the turbulent cascade model \citep{Howes:2008b} 
to calculate the total proton and electron heating resulting from
collisionless damping of the electromagnetic fluctuations of the
\Alfvenic turbulent cascade. The model employs the linear
collisionless gyrokinetic dispersion relation \citep{Howes:2006} to
determine both the linear kinetic damping rate via the Landau
resonances and the nonlinear energy cascade rate, so the resulting
heating prescription is only valid in the gyrokinetic limit, $\omega
\ll \Omega_p$.

The resulting prescription for the ratio of proton-to-electron heating
for $T_p/T_e>1$ is given by
\begin{equation}
Q_p/Q_e = c_1\frac{c_2^2 + \beta_p^\alpha}{c_3^2 + \beta_p^\alpha}
\sqrt{\frac{m_p T_p}{m_e T_e}} e^{-1/\beta_p}
\label{eq:fit}
\end{equation}
where $c_1=0.92$, $c_2=1.6/(T_p/T_e)$, $c_3=18+5 \log (T_p/T_e)$, and
$\alpha=2-0.2 \log (T_p/T_e)$. The requirement that the proton
cyclotron resonance plays a negligible role in the dynamics and
dissipation of the turbulence enables the regime of validity of this
turbulent heating prescription to be quantified.  The limit of the
regime of validity can be expressed as a constraint on the 
minimum width of the inertial range $(k_{0}\rho_p)^{-1}_{min}$, shown
in panel (a) of \figref{fig:kx0} as a contour plot in the
$(\beta_p,T_p/T_e)$ parameter space.

\begin{figure}
\resizebox{3.1in}{!}{\includegraphics{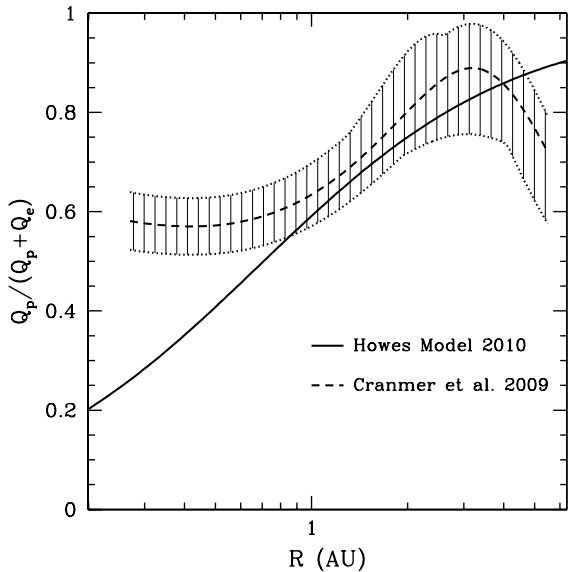}}
\caption{\label{fig:qpqtot} A comparison of  
the proton-to-total heating rate $Q_p/(Q_p+Q_e)$ as a function of
heliocentric radius $R$: the solid line is the prediction based on
the turbulent heating prescription given by \eqref{eq:fit}
\citep{Howes:2010d}, and the dashed line is the empirical estimate of
\citet{Cranmer:2009}, where dotted lines indicate an error estimate
to account for modeling and observational uncertainties. }
\end{figure}

Substituting the values of $\beta_p(R)$ and $T_p/T_e(R)$ specified in
\secref{sec:model} into \eqref{eq:fit} enables the calculation of the predicted 
ratio of proton-to-electron turbulent heating $Q_p/Q_e (R)$ as a
function of heliocentric radius. In \figref{fig:qpqtot}, we plot the
predicted ratio of proton-to-total turbulent heating $Q_p/(Q_p+Q_e)$
vs.~heliocentric radius $R$.

We compare this theoretical prediction to the the empirical estimate
by \citet{Cranmer:2009} (dashed) for $0.8$~AU $\lesssim R\le 5.4$~AU
in \figref{fig:qpqtot}. The error estimates (dotted) in this figure
are derived by attempting to account for the error arising from both
modeling and observational uncertainties.  Modeling uncertainties are
derived by taking the curves for outflow speeds of 650, 700, 750, and
800 km s$^{-1}$ in Figure~3(a) and colatitudes of 0, 15, and 30 degrees in
Figure~4(b) of \citet{Cranmer:2009}.  The observational uncertainties
in the calculated turbulent heating due to observed variations in the
proton and electron temperatures and electron heat flux are estimated
roughly by taking $\pm10$\% of the proton-to-total turbulent heating
ratio.  The error estimates (dotted) plotted in \figref{fig:qpqtot}
are determined by taking the outer envelope of these modeling and
observational uncertainties.

We find generally good agreement between the prediction of the
turbulent heating prescription (solid) and the empirical estimate
by \citet{Cranmer:2009} (dashed) for $0.8$~AU$ \lesssim R\le
5.4$~AU.  The disagreement at $R\lesssim 0.8$~AU, as we shall see in
\secref{sec:limits}, is attributed to the violation of the gyrokinetic
approximation, and, therefore, to exceeding the limits of validity of
the turbulent heating prescription. The downturn in the empirical
estimate of $Q_p/(Q_p+Q_e)$ (dashed) seen at $R > 3$~AU may be an
artifact of the bifurcation of the electron temperatures
measured by \emph{Ulysses}, as seen in Figure~1(a) of 
\citet{Cranmer:2009},  and may not represent an actual decrease in the
proton-to-total turbulent heating ratio for the high-speed wind.
\citet{Cranmer:2009} noted that this appeared to be a solar cycle effect,
but further work will required to ascertain the significance of this
downturn.

We can compare our prediction of the turbulent heating based on
\eqref{eq:fit} with predictions based on simple theoretical models
of the turbulent heating by \citet{Cranmer:2009}, presented in
Figure~5 of their paper. Their approach used a quasilinear framework
to estimate the proton-to-electron heating rates for three particular
models of the distribution of turbulent energy in wavevector space: an
isotropic distribution, a slab distribution of only parallel
wavevectors, and a ``two-dimensional'' (2D) distribution of nearly
perpendicular wavevectors. All three of the models showed significant
disagreement with the empirically determined heating ratio.  The slab
model predicted 100\% proton heating, the isotropic model
overestimated the proton heating for $R\lesssim 2$~AU, and the 2D
model significantly underestimated the proton heating at $R\gtrsim
1$~AU. In light of these results, the agreement between the
proton-to-total turbulent heating ratio predicted by \eqref{eq:fit}
and the empirical estimate in \figref{fig:qpqtot} is quite good.  This result suggests that
the turbulent cascade model
\citep{Howes:2008b} captures the dominant physical mechanisms
(described qualitatively in \secref{sec:turbmodel}) that play a role
in the dissipation of solar wind turbulence at $\gtrsim 0.8$~AU.

The disagreement between the prediction of the proton-to-total
turbulent heating ratio (solid) and the empirical estimate (dashed) in
\figref{fig:qpqtot} can be understood if we evaluate the limit of 
the regime of validity of the turbulent heating prescription
(eq.~[\ref{eq:fit}]) using the plasma parameters specified for the solar
wind model in
\secref{sec:model}.  The violation of the gyrokinetic approximation,
or, equivalently, the point at which the proton cyclotron resonance
begins to play a non-negligible role in the dynamics and dissipation
of the turbulence, is cast as a requirement for the minimum dynamic
range spanned by the inertial range, given by the driving scale
divided by the proton Larmor radius, $L_0/\rho_p \sim (k_{0}
\rho_p)^{-1}$. We denote this as the  minimum ``width'' of the 
inertial range, $(k_{0} \rho_p)^{-1}_{min}$, plotted in panel (a) of
\figref{fig:kx0} as a contour plot in the $(\beta_p,T_p/T_e)$
parameter space.  The proton Larmor radius $\rho_p = v_{tp}/\Omega_p
=c (2 T_p m_p)^{1/2}/(q_p B)$ is easily determined as a function of
heliocentric distance $R$ given the models for $T_p(R)$ and $B(R)$
given in \secref{sec:model}. Estimating isotropic driving wavenumber
of the turbulence $k_0(R)$, however, requires the incorporation of
additional empirical constraints.

\subsection{Evolution of the Width of the Inertial Range}
\label{sec:width}
We interpret the isotropic driving scale $L_0 = 2 \pi/k_0$ to be the
outer scale of the inertial range, and we identify this scale
observationally as the break in the solar wind magnetic energy
spectrum from the $f^{-1}$ energy containing range to the $f^{-5/3}$
inertial range. To estimate $k_0$, we employ measurements of magnetic
energy spectrum from \emph{Helios 2} data published in Figure~23 of
\citet{Bruno:2005}.  The frequency $f_0$ of the spectral break
marking the outer scale of the inertial range measured from this
figure is given in Table~\ref{tab:k0}. Each of these spectra are
measured in the same corotating fast stream with a velocity $v_{sw}
\simeq 700$~km s$^{-1}$ (taken from Figure~17 of \citet{Bruno:2005}), so we
may calculate the corresponding inertial range outer scale length
$L_0=v_{sw}/f_0$, or wavenumber $k_0 =2 \pi/L_0$.
\begin{table}
\begin{center}
\begin{tabular}{|c|c||c|c|}
\hline
$R$ (AU) & $f_0$ (HZ) & $L_0$ (km) & $k_0$ (rad km$^{-1}$)\\
\hline
0.3 & $6\times 10^{-3}$ & $1.2 \times 10^{5}$ & $5.2 \times 10^{-5}$
\\ 0.7 &$1.5\times 10^{-3}$ & $4.7 \times 10^{5}$ & $1.3 \times
10^{-5}$ \\ 0.9 &$6\times 10^{-4}$ & $1.2 \times 10^{6}$ & $5.2 \times
10^{-6}$ \\
\hline
\end{tabular}
\end{center}
\caption{Measured values of the frequency $f_0$ of the outer scale of 
the inertial range of solar wind turbulence measured by \emph{Helios~2}.
\label{tab:k0} 
}
\end{table}

The function 
\begin{equation}
k_0 = K_{0} \left(\frac{R}{1\mbox{ AU}}\right)^{-2}
\label{eq:k0}
\end{equation}
provides a reasonable fit for the evolution of $k_0$ as a function of
heliocentric radius $R$ with the value $ K_{0}= 5 \times
10^{-6}$~rad/km. Although this fit is based solely on measurements in
the inner heliosphere and may not be an accurate representation of the
evolution of $k_0(R)$ for $R>0.9$~AU, we will see that
 error in the estimation of $k_0$ at $R\gtrsim 1$~AU does not strongly 
impact the applicability of the turbulent heating prescription 
for the plasma parameters derived from this solar wind model.

\subsection{Evaluation of the Limits of Validity of the Turbulent Heating Prescription}
\label{sec:limits}

\begin{figure}
\resizebox{3.1in}{!}{\includegraphics{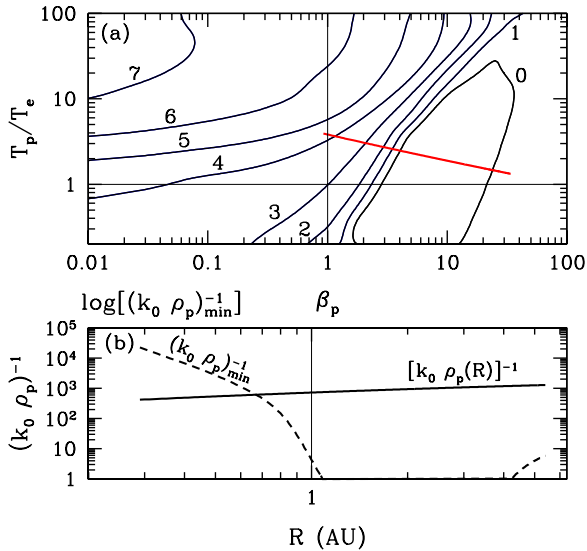}}
\caption{\label{fig:kx0} (a) Logarithmic contour plot of the  minimum width of the 
inertial range $(k_{0} \rho_p)^{-1}_{min}$ over the plane $(\beta_p,
T_p/T_e)$, the condition required for the the proton cyclotron
resonance to play a negligible role. The red line denotes the
evolution plasma parameters of the solar wind model in
\secref{sec:model} from $R=0.29$~AU (left end) to $R=5.4$~AU (right
end).  (b) The minimum width of the inertial range $(k_{0}
\rho_p)^{-1}_{min}$ along the red line in panel (a) (dashed)
vs.~the estimated width of the inertial range $(k_{0} \rho_p)^{-1}$
for the solar wind model considered here (solid).  For the heating
prescription to be valid, the dashed line must fall below the solid
line.  }
\end{figure}

Using the values of $\rho_p(R)$ and $k_0(R)$ derived here, we
calculate the width of the inertial range $k_{0} \rho_p$ as a function
of heliocentric radius and compare it to the constraint on $(k_{0}
\rho_p)^{-1}_{min}$, as shown in \figref{fig:kx0}. Panel (a)
presents a logarithmic contour plot of the constraint $(k_{0}
\rho_p)^{-1}_{min}$ over the $(\beta_p,T_p/T_e)$ parameter space.  Also shown is
the path though this parameter space (red) traversed by the solar wind
model from 0.29~AU to 5.4~AU. In panel (b), the value of the
width of the inertial range $(k_{0} \rho_p)^{-1}$ for our solar wind
model (solid) is plotted against the constraint on the minimum width
of the inertial range $(k_{0} \rho_p)^{-1}_{min}$ for the validity of
our heating model (dashed). (For the heating prescription to be valid,
the dashed line must fall below the solid line in this plot.)  As
previously mentioned, for the plasma parameters $\beta_p$ and
$T_p/T_e$ specified by the solar wind model in \secref{sec:model}, the
constraint on $(k_{0} \rho_p)^{-1}_{min}$ does not strongly restrict the
applicability of the heating prescription for $R>1$~AU.  It is clear,
however, that this constraint is violated for heliocentric distances
$R \lesssim 0.7$~AU, signaling that the gyrokinetic representation of
the turbulent dissipation mechanisms is no longer valid because the
proton cyclotron resonance has begun to play a non-negligible
role. Significantly, this limit to the validity of our prediction for
the proton-to-total turbulent heating ratio coincides with the point
where our prediction (solid) begins to fall significantly below the
empirical estimate (dashed) in
\figref{fig:qpqtot}. If the contribution to proton heating from the
proton cyclotron resonance becomes non-negligible for heliocentric
distances $R \lesssim 0.8$~AU, a physical process not represented in
the gyrokinetic turbulent heating prescription, then the result would
be an empirically estimated proton-to-total heating rate that exceeds
the gyrokinetic predictions at these radii, in agreement with the
behavior shown in \figref{fig:qpqtot}.

\section{Discussion}
The results presented in \secref{sec:results} suggest the following
consistent picture of the physical mechanisms in the high-speed solar
wind that are responsible for the dissipation of the turbulence and
that lead to heating of the plasma protons and electrons.

In the inner heliosphere at $R\lesssim 0.8$~AU, the typically high
$T_p/T_e$ and low $\beta_p$ conditions (coupled with a slightly
smaller width of the inertial range $(k_0 \rho_p)^{-1}$; see panel (b)
of \figref{fig:kx0}) in high-speed streams lead to a turbulent cascade
in which the small scale turbulent fluctuations can reach the proton
cyclotron frequency before the turbulence is collisionlessly damped
via the Landau resonances. Therefore, one may expect to observe
greater heating of the protons than that predicted by the turbulent
heating prescription given by \eqref{eq:fit}, a result based on a
gyrokinetic cascade model \citep{Howes:2008b}.  By the time the
turbulence has reached $R > 0.8$~AU, the decrease of $T_p/T_e$ and
increase of $\beta_p$ (coupled with a slight increase in the width of
the inertial range $(k_0 \rho_p)^{-1}$; see panel (b) of
\figref{fig:kx0}), lead to plasma conditions in which the turbulent
cascade no longer is affected by the proton cyclotron resonance before
it is terminated by collisionless damping via the Landau
resonances. Thus, for the range $0.8$~AU $\lesssim R\le 5.4$~AU, the
turbulent heating prescription in the gyrokinetic limit adequately
represents all of the physical mechanisms needed to reproduce the
empirically estimated proton-to-total turbulent heating ratio, as seen
in \figref{fig:qpqtot}.

This theoretical prediction of non-negligible proton cyclotron damping
within the inner heliosphere at $R \lesssim 0.8$~AU is consistent with
observational evidence. Proton cyclotron damping is expected to lead
to heating of the protons in the direction perpendicular to the local
mean magnetic field \citep{Lehe:2009}. In the absence of such
perpendicular proton heating, double adiabatic evolution would lead to
a constant value of $T_{\perp p}/B$ as a function of heliocentric
radius \citep{Chew:1956}.  \emph{Helios} observations demonstrate
that the constancy of $T_{\perp p}/B$ is indeed violated within the
inner heliosphere \citep{Marsch:1983}.

It is important to note that the observed non-adiabatic $T_{\perp
p}/B$ does not directly identify the physical mechanism
responsible. In addition to heating via the proton cyclotron
resonance, several other physical mechanisms could lead to the
observed behavior, including stochastic proton heating and kinetic
proton temperature anisotropy instabilities. 

Stochastic heating had been proposed as a mechanism for perpendicular
proton heating for some time
\citep{Johnson:2001,Chen:2001,White:2002,Voitenko:2004,Bourouaine:2008},
and recently \citet{Chandran:2010a} have put forth strong theoretical
and numerical evidence for stochastic perpendicular heating of protons
at low plasma $\beta$ by kinetic \Alfven waves of sufficient
amplitude. \citet{Chandran:2010b} has used these results for
stochastic heating to explain proton and minor ion perpendicular
temperature observations in coronal holes. For the low plasma $\beta$
conditions found in the inner heliosphere, it is possible that this
mechanism of stochastic heating could explain the perpendicular proton
temperature measurements \citep{Marsch:1983} and the empirically
estimated proton heating \citep{Cranmer:2009} at $R
\lesssim 0.8$~AU.  

In addition, kinetic proton temperature anisotropy instabilities in
the spherically expanding solar wind flow have been shown to play a
role in regulating the proton temperature anisotropy
\citep{Kasper:2002,Hellinger:2006,Bale:2009}.  Although these
instabilities cannot lead to a net heating of the proton species, they
can mediate a transfer of energy from the parallel to the
perpendicular temperature, and vice versa, and so may be a cause of
the observed deviation from  double adiabatic evolution of the
perpendicular temperature of the protons \citep{Marsch:1983}.

There are several possible limitations of the application of the
turbulent heating prescription given by \eqref{eq:fit} to the problem
of the heating due to the dissipation of solar wind turbulence. First,
the model does not account for energy in compressible wave modes, such
as the collisionless manifestation of the fast and slow MHD wave
modes. If significant energy exists in these compressible modes, any
heating due to the dissipation of these modes must be handled
separately.  Second, the cascade model is constructed specifically for
the case of balanced \Alfven wave energy fluxes up and down the local
magnetic field, corresponding to zero cross helicity.  Observations of
the cross helicity in the high-latitude wind typically show non-zero
normalized cross helicities varying over the range $0.2 \le
\sigma_c\le 0.6$ \citep{Bavassano:2000a,Bavassano:2000b}. Note,
however, that a cross helicity of $\sigma_c=0.6$ corresponds to the
amplitude of anti-sunward waves only a factor of 2 larger than the
sunward waves. Since the typical energy cascade rates in strong MHD
turbulence vary linearly with the wave amplitudes
\citep{Goldreich:1995,Howes:2008b}, this level of imbalance in the turbulence 
is unlikely to yield significant qualitative differences compared to
the balanced case. Third, the simplifying assumption of a fully
ionized proton and electron plasma with isotropic Maxwellian
equilibrium velocity distributions neglects the physical variations
that may arise from the more complicated equilibrium conditions often
observed in the solar wind.  Such conditions include temperature
anisotropy with respect to the local magnetic field direction (often
treated using a bi-Maxwellian equilibrium distribution), significant
deviations from a Maxwellian distribution at high energy, and the
presence of minor ions, particularly helium. Nonetheless, we believe
that the results presented here represent a significant step forward
in our understanding of the mechanisms responsible for proton and
electron heating in the turbulent solar wind. Future work will explore
the implications of these additional effects if any of them appears to
impact significantly the findings presented here.

\section{Conclusions}
In the effort to identify the physical mechanisms that govern the
dissipation of solar wind turbulence and lead to heating of the solar
wind protons and electrons, \citet{Cranmer:2009} made a great stride
forward by determining an empirical estimate of the proton-to-total
plasma heating in the high-speed solar wind using \emph{Helios} and
\emph{Ulysses} data. Based on a turbulent energy cascade  
model for low-frequency, anisotropic \Alfvenic turbulence in a weakly
collisional plasma \citep{Howes:2008b}, \citet{Howes:2010d}
constructed an analytical prescription for the total
proton-to-electron heating resulting from collisionless damping of the
electromagnetic fluctuations of the \Alfvenic turbulence.  

Applying this turbulent heating prescription to predict the
proton-to-total plasma heating in the high-speed solar wind, we obtain
the following results, as shown in \figref{fig:qpqtot}: (1) the
prediction agrees well with the empirical estimate for
$0.8$~AU~$\lesssim R \le 5.4$~AU; (2) the predicted proton heating
falls below the empirical estimate for $R \lesssim0.8$~AU.
Investigating the cause of the disagreement for $R \lesssim 0.8$~AU,
we see, in  panel (b) of \figref{fig:kx0}, that the turbulent
heating prescription in the gyrokinetic limit ceases to be valid for
$R \lesssim 0.7$~AU. This failure of the prescription's validity has
physical meaning. In this region, the cascade model predicts that the
turbulent fluctuations will reach the proton cyclotron frequency
before they are damped via the Landau resonances.  Therefore, we
expect that proton cyclotron damping will cause additional proton
heating, leading to an underestimate of the proton-to-total heating
ratio, as seen in \figref{fig:qpqtot}.

These results suggest the following physical picture of the turbulent
cascade and plasma heating in the high-speed solar wind.  In the
inner heliosphere at $R \lesssim 0.8$~AU, the turbulent cascade can
reach the proton cyclotron frequency, leading to a level of proton
heating higher than that predicted by the turbulent heating
prescription in the gyrokinetic limit. But for heliocentric radii $R
\gtrsim 0.8$~AU, collisionless damping via the Landau resonances
terminates the turbulent cascade before the proton cyclotron resonance
is reached, so the turbulent heating prescription in the gyrokinetic
limit contains all of the necessary physical mechanisms needed to
reproduce the empirically estimated proton-to-total heating ratio.

%

\acknowledgments

G.~G.~H. thanks Steve Cranmer for insightful discussions and for
providing his empirical heating data with error estimates. The work
has been supported by NSF CAREER Award AGS-1054061 and NASA
NNX10AC91G.


\clearpage

\end{document}